\documentstyle[12pt]{article}
\newcommand{\nc}{\newcommand}
\nc{\al}{\alpha}
\nc{\be}{\beta}
\nc{\ga}{\gamma}
\nc{\de}{\delta}
\nc{\ep}{\epsilon}
\nc{\ze}{\zeta}
\nc{\la}{\lambda}
\nc{\rh}{\rho}
\nc{\si}{\sigma}
\nc{\Ga}{\Gamma}
\nc{\Si}{\Sigma}
\nc{\ptl}{\partial}

\title{The Low Energy Effective Lagrangian for Photon Interactions in Any
       Dimension}
\author{A. Ritz and R. Delbourgo\\
        $\;$ \\
        Department of Physics, University of Tasmania \\
        GPO Box 252C Hobart, Tasmania, Australia 7001.}
\date{}

\begin{document}

\maketitle

\begin{abstract}
\thispagestyle{plain}
 The subject of low energy photon-photon scattering is considered in
arbitrary dimensional space-time and the interaction is widened to include
scattering events involving an arbitrary number of photons. The effective
interaction Lagrangian for these processes in QED has been determined in a
manifestly invariant form. This generalisation resolves the structure of
the weak-field Euler-Heisenberg Lagrangian and indicates that the
component invariant functions have coefficients related, not only to the
space-time dimension, but also to the coefficients of the Bernoulli
polynomial.
\end{abstract}

\section{Introduction}

That an interaction between electromagnetic fields and the vacuum of the
Dirac field would allow such nonlinear processes as the scattering of
photons by photons has been known for some time. Even before the emergence
of the full scale theory of
quantum electrodynamics (QED) Halpern, Euler, and Heisenberg, among others,
discussed the phenomenon in the mid 1930s \cite{halp34,eul35,eul36,heis36}.
However, it was in the early nineteen fifties that the subject was
considered in detail using QED, and Karplus \& Neuman
\cite{karplus50,karplus51} determined the full photon-photon
scattering cross-section.

The work presented in this paper is related to a desire to generalise the
discussion of photon interactions to arbitrary dimensional space-time and
to consider how the parameters of the interaction change in this
case, and also with the number of scattering photons. This will
be done by investigating the underlying {\em effective Lagrangian\/} for the
interaction, and little consideration will be given to
measureable cross-sections. The reason for this being that the cross-section
determined by Karplus \& Neuman, and others subsequently \cite{beres82},
for four photon scattering in four dimensional space-time is extremely small
(i.e. $\si \sim 10^{-30} $cm$^2$). This is perhaps not unexpected for such
an interaction, but it does mean that obtaining experimentally testable
values is of limited importance at this stage. The
generalisation to arbitrary dimensional space-time serves a twofold
purpose. Firstly, it is useful to know the form of the result in the
context of dimensional regularisation. More
importantly perhaps it will provide greater insight into the structure of
the Lagrangian in the dimensions of practical interest, i.e. $D=2,3$ and
$4$. As a scattering process involving the fewest number of photons
(i.e. four) will have the largest cross-section, generalisation to
interactions involving an
arbitrary number of photons will also primarily be of interest in further
resolving the structure in the Lagrangian.

A scattering process involving $N$ photons can be represented by a sum
over permutations of an $N$-photon electron loop. i.e. a coupling of $N$
photons via an intermediate loop of electron and positron propagators.
Direct evaluation of the Feynman amplitude in this case would be quite
involved and this
problem is most easily tackled by taking, as a starting point, the full
1-loop Lagrangian for the field involving a summation of all possible
couplings between the electromagnetic field and a virtual matter field. For
reasons of
clarity it is useful to split the discussion of $N$-photons interactions
into two cases: those involving an even number of photons; and those
involving an odd number. We shall proceed now to consider the first of
these.

\section{2N-Photon Case}
Interactions involving an even number of photons particularly lend
themselves to the term
photon-photon scattering, as a situation involving the same number of
initial and final photons can easily be envisaged. In the discussion which
follows we shall implicitly assume that the interactions involve $2N$
photons, where $N$ is integral, and thus the number of interaction photons
involved is always even. In order to discuss
these cases we can recall the 1-loop electromagnetic field Lagrangians
obtained by Brown \& Duff \cite{brown75} for the cases of
bosonic and fermionic virtual particle loops:
\begin{eqnarray}
 {\cal L}_I^{(0)} & = & \frac{1}{(4\pi)^{D/2}} \int_0^{\infty}
                        \frac{\mbox{d}s}{s^{1+D/2}}e^{-m^2 s}
  \exp\left[-\frac{1}{2} {\rm Tr}\; \ln \left(\frac{\sin eFs}{eFs}\right)
\right],
             \label{spin0}\\
 {\cal L}_I^{(1/2)} & = & - \frac{1}{2(4\pi)^{D/2}} \int_0^{\infty}
                        \frac{\mbox{d}s}{s^{1+D/2}}e^{-m^2 s}
            {\rm Tr}\; \exp\left[\frac{1}{2}es \si \cdot F \right] \nonumber\\
	       &   & \;\;\;\;\;\;\;\;\;\;\;\;\;\;
  \exp\left[-\frac{1}{2} {\rm Tr}\; \ln \left(\frac{\sin eFs}{eFs}\right)
\right]
                       . \label{spin1/2}
\end{eqnarray}
These results being generalisations of the four-dimensional Lagrangians
obtained by Schwinger \cite{schwinger51,schwinger73}. Brown \& Duff used
momentum space functional methods to consider the Lagrangian in
any field theory due to 1-loop quantum vacuum effects. The reader is
directed to their paper for details of the derivation, while we shall
simply use their results to consider the case of $2N$-photon scattering.
It should be noted that these expressions are valid in strong
fields as they contain all orders of the coupling to the electromagnetic
field. Taking only the lowest order terms will correspond to the
situation of weak fields.

As it is a simple starting point we shall first consider the situation of
photon scattering via a virtual meson loop (i.e. spin zero propagator
particles). Techniques used in this case will then be applied to the
slightly more complex process in which photon scattering is mediated by an
electron/positron loop.

\subsection{Scalar Case}
Our starting point here will be the 1-loop Lagrangian of Brown \& Duff
(Eq. \ref{spin0}) which can be written in the following form
\begin{eqnarray}
 {\cal L}_I^{(0)} & = & \frac{1}{(4\pi)^{D/2}} \int_0^{\infty}
                        \frac{\mbox{d}s}{s^{1+D/2}}e^{-m^2 s}
  \exp\left[\frac{1}{2}{\rm Tr}\;\ln f(iF)\right]  \label{s0}
\end{eqnarray}
where
\begin{eqnarray*}
 f(z) & = & \frac{esz}{\sinh(esz)}.
\end{eqnarray*}
This Lagrangian, although it must be Lorentz  covariant, is not
explicitly so. Our task will be to extract a given order of the
interaction (i.e. $2N$-photon scattering) and express this part of the
Lagrangian in a manifestly covariant form.
To this end we can follow the procedure
described by Delbourgo \& Matsuki \cite{delbourgo85} and
suppose that the function
$f(z)$ possesses a series expansion of the form
\begin{eqnarray}
 f(z) & = & 1 + \sum_{n=1}^{\infty} b_n x^{2n}, \label{f}
\end{eqnarray}
which is clearly true in the case above. Then writing the exponential and
logarithmic functions as power series we can obtain
\begin{eqnarray*}
 {\cal L}_I^{(0)} & = & \frac{1}{(4\pi)^{D/2}} \int_0^{\infty}
                        \frac{\mbox{d}s}{s^{1+D/2}}e^{-m^2 s}
      \sum_{l=0}^{\infty} \frac{1}{l!} \left[{\rm Tr}\; \sum_{m=1}^{\infty}
              a_m F^{2m}\right]^l,
\end{eqnarray*}
where new constants $a_m$ have been introduced to simplify the
expression. As this relation is now in the form of a single power
series expansion we can pick out any given term, and in particular the
general term, which corresponds to a $2N$-photon
scattering interaction, will be given by
\begin{eqnarray}
 {\cal L}_I^{(0)} & = & \frac{1}{(4\pi)^{D/2}} \int_0^{\infty}
                        \frac{\mbox{d}s}{s^{1+D/2}}e^{-m^2 s}
    \sum_{\{n_i\}}^{n_1+2n_2+\cdots+Nn_N=N} \frac{1}{n_1!n_2!\ldots n_N!}
              \nonumber\\
  & & \;\;\; \times(a_1 {\rm Tr}\;F^2)^{n_1}(a_2 {\rm Tr}\;F^4)^{n_2}\cdots
                 (a_N {\rm Tr}\;F^{2N})^{n_N}. \label{covl}
\end{eqnarray}
In this expression the summation over $n_i$ runs from zero to a certain
integer such that the partitioning constraint,
\begin{eqnarray*}
 n_1+2n_2+3n_3+\cdots+Nn_N & = &N,
\end{eqnarray*}
is satisfied. We should note here that implicit in this selection of a
particular order of the expansion as a valid Lagrangian is any
renormalisation of charge, and corresponding scale changes for the
fields, which may be needed.

This Lagrangian is now in a manifestly invariant form and the problem has
reduced to determining the coefficients $a_m$. Fortunately a method for
determining these factors has been developed in a rather different area
of field theory. Delbourgo \& Matsuki \cite{delbourgo85}, when considering
gravitational and chiral anomalies \cite{alvarez83,endo85} obtained
the following expression in terms of a two-form matrix version of the
Riemann tensor $R$
\begin{eqnarray*}
 A_D & = & C_D \exp \left[ \frac{1}{2} {\rm Tr}\; \ln f(iR) \right].
\end{eqnarray*}
As we see this is identical to the second exponential term in the 1-loop
Lagrangian (Eq. \ref{s0}) apart from the use of $R$ in place of $F$. The
fact that the same mathematics has turned up in apparently unrelated
areas of field theory is certainly of note, but
leaving this interesting point aside we can use the following expression,
obtained by Delbourgo \& Matsuki \cite{delbourgo85},
to determine the coefficients $a_m$.
\begin{eqnarray*}
 a_m & = & \frac{i(-1)^{m+1}}{8\pi m} \oint \mbox{d}z \; z^{-2m}
          \frac{\mbox{d}}{\mbox{d}z} \ln f(z).
\end{eqnarray*}

This general result can be applied to the particular case of the spin-0
Lagrangian in which we have
\begin{eqnarray*}
  f(z) & = & \frac{esz}{\sinh (esz)}.
\end{eqnarray*}
Determination of the coefficients can then be achieved by performing the
integration and noting that
\begin{eqnarray*}
 \frac{z}{e^z -1} & = & \sum_{n=0}^{\infty} \frac{B_n}{n!}z^n,
\end{eqnarray*}
where $B_n$ are the Bernoulli numbers. We obtain
\begin{eqnarray*}
 a_m & = & \frac{(-1)^{m+1}(2es)^{2m}}{4m} \frac{B_{2m}}{(2m)!}.
\end{eqnarray*}

Thus, for a $2N$-photon process mediated by a virtual meson
loop, the interaction Lagrangian is given by
\begin{eqnarray*}
 {\cal L}_I^{(0)} & = & \frac{1}{(4\pi)^{D/2}} \int_0^{\infty}
                        \frac{\mbox{d}s}{s^{1+D/2}}e^{-m^2 s}e^{2N}s^{2N}
    \sum_{\{n_i\}}^{n_1+2n_2+\cdots+Nn_N=N} \frac{1}{n_1!n_2!\ldots n_N!}\\
  & & \;\;\; \times(a'_1 {\rm Tr}\;F^2)^{n_1}(a'_2 {\rm Tr}\;F^4)^{n_2}\cdots
                 (a'_N {\rm Tr}\;F^{2N})^{n_N},
\end{eqnarray*}
where
\begin{eqnarray*}
 a'_m & = & \frac{(-1)^{m+1}2^{2m}}{4m} \frac{B_{2m}}{(2m)!}.
\end{eqnarray*}
We can now proceed to perform the integration, obtaining
\begin{eqnarray*}
 {\cal L}_I^{(0)} & = & \frac{1}{(4\pi)^{D/2}}\frac{e^{2N}}{(m^2)^{2N-D/2}}
       \int_0^{\infty} (m^2s)^{(2N-D/2)-1} e^{-m^2 s}\mbox{d}(m^2 s) \\
 & &  \sum_{\{n_i\}}^{n_1+2n_2+\cdots+Nn_N=N} \frac{1}{n_1!n_2!\ldots n_N!}
         (a'_1 {\rm Tr}\;F^2)^{n_1}(a'_2 {\rm Tr}\;F^4)^{n_2}\cdots(a'_N
              {\rm Tr}\;F^{2N})^{n_N}.
\end{eqnarray*}
The integral in this expression is simply the defining relation for the
$\Ga$-function, i.e.
\begin{eqnarray*}
  \Ga (x) & = & \int_0^{\infty} t^{x-1} e^{-t} \mbox{d} t,
\end{eqnarray*}
and thus
\begin{eqnarray*}
 {\cal L}_I^{(0)} & = & \frac{(e^2)^N}{(4\pi)^{D/2}}\frac{1}{(m^2)^{2N-D/2}}
      \Ga(2N-D/2)\sum_{\{n_i\}}^{n_1+2n_2+\cdots+Nn_N=N}
          \frac{1}{n_1!n_2!\ldots n_N!} \\
 & & \;\;\;\;\;\times(a'_1 {\rm Tr}\;F^2)^{n_1}(a'_2 {\rm Tr}\;F^4)^{n_2}\cdots
        (a'_N {\rm Tr}\;F^{2N})^{n_N}.
\end{eqnarray*}

We will find that it is particularly illuminating in the spinor case to use
the the coefficients of the Bernoulli polynomial $B_n(x)$ \cite{prudnikov},
\[ \frac{ze^{xz}}{e^z-1} = \sum^{\infty}_{n=0} B_n(x) \frac{z^n}{n!},\]
in place of the Bernoulli numbers and for later comparison we shall
introduce them here. With the substitution
$ B_n(0) = B_n $ the invariant coefficients become
\[ a'_m = \frac{(-1)^{m+1} 2^{2m}}{4m (2m)!} B_{2m}(0),\]
and we can write the effective interaction Lagrangian in the following
compact form
\begin{eqnarray}
 {\cal L}_I^{(0)} & = & \frac{(e^2)^N}{(4\pi)^{D/2}}\frac{1}{(m^2)^{2N-D/2}}
      \Ga(2N-D/2) \sum_{\{n_i\}}^{n_1+2n_2+\cdots+Nn_N=N}
                 \frac{1}{n_1!n_2!\ldots n_N!} \nonumber \\
 & & \times\left(\frac{B_2(0)}{2} {\rm Tr}\;F^2\right)^{n_1}
     \left(-\frac{B_4(0)}{12} {\rm Tr}\;F^4\right)^{n_2} \nonumber\\
 & & \;\;\;\;\;\;\;\;\;\;\;\;\;\;\;\;    \cdots
     \left(\frac{(-1)^{N+1} 2^{2N}}{4N (2N)!}B_{2N}(0)\; {\rm
Tr}\;F^{2N}\right)^{n_N}. \label{sp0}
\end{eqnarray}

This is our final expression for the Lorentz invariant low energy effective
interaction Lagrangian for
the electromagnetic field due to spin-0 quantum loop effects.
Although this superficially describes all $2N$-photon scattering
interactions in $D$ dimensions, we should note that in
some cases the $\Ga$-function may be undefined, i.e. for large even $D$
and small $N$. This is not a serious problem as in the cases of most
interest (e.g. 4 photon scattering in $D=4$) the result will be either
finite or manageable via regularisation.

As a particular example we can consider the case of four photon
scattering ($N=2$) in four dimensional space-time. In this case the
constraint over the $\{n_i\}$ implies that the only two terms allowed
will have $n_1=2$  and $n_2=1$. Thus we immediately recover the
electromagnetic field invariants
\begin{eqnarray*}
 ({\rm Tr}\;F^2)^2, \;\;\;\; \mbox{and} \;\;\;\; {\rm Tr}\;F^4,
\end{eqnarray*}
obtained in previous discussions of four photon scattering. The
coefficients can be evaluated easily from the general expression,
and in this case we obtain
\begin{eqnarray*}
 {\cal L}_I^{(0)} & = & \frac{\al^2}{90m^4}
         \left[\frac{5}{16}({\rm Tr}\;F^2)^2 + \frac{1}{4}{\rm Tr}\;F^4\right],
\end{eqnarray*}
which is precisely the low energy limit of the result
obtained by Schwinger \cite{schwinger51,schwinger73}.

\subsection{Spinor Case}
For this case our starting point is the spin-1/2 1-loop Lagrangian of
Brown \& Duff (Eq. \ref{spin1/2})
where we see the spinor case introduces the extra trace over spinor indices,
\begin{eqnarray*}
  {\rm Tr}\; \exp\left[\frac{1}{2}es \si \cdot F \right].
\end{eqnarray*}
Before proceeding to use the techniques developed in the scalar case this
trace must be evaluated and in order to do so it is neccessary to consider
the cases of odd and even dimensional space-times separately.

\subsubsection{Even $D$}
Evaluation of the spinor trace is essentially a purely mathematical
exercise, and in order to simplify the process as much as possible we
can perform the calculation in Euclidean rather than Minkowskian space.
In this case the metric, $g_{\mu\nu}$, assumes the form of the
identity matrix
and for even dimensions the antisymmetric electromagnetic field tensor
can be written in the following canonical form,
\begin{eqnarray*}
  (F)^{\mu\nu} & = & \left(\begin{array}{ccccccc}
      0       & \la_1    & 0      &          &\cdots &       &          \\
      -\la_1  & 0        &        &          &       &       &          \\
        0     &          & 0      & \la_2    &       &       &          \\
              &          & -\la_2 & 0        &       &       &          \\
      \vdots  &          &        &          &\ddots &       &          \\
              &          &        &          &       & 0     & \la_{D/2}\\
              &          &        &          &       &-\la_{D/2} & 0
                \end{array} \right),
\end{eqnarray*}
where $\la_i$ are the eigenvalues. This canonical form can be achieved by a
similarity
transformation, which will not effect the commutation relations between
the Lorentz group generators, but corresponds to choosing some canonical
basis for the $\ga$-matrices.

If we consider a coupling of the electromagnetic field to a
spin-1/2 field via
\begin{eqnarray*}
 \si_{\mu\nu} & = & \frac{i}{2}\left[\ga_{\mu},\ga_{\nu}\right] =
     \frac{i}{2} \left(\ga_{\mu}\ga_{\nu}-\ga_{\nu}\ga_{\mu} \right),
\end{eqnarray*}
then this will result in the invariant matrix $\si \cdot F$ which we can
represent as
\begin{eqnarray*}
 \frac{1}{2} \si F & = & \frac{1}{2}i\ga_{\mu}\ga_{\nu}F^{\mu\nu}
     =  i \left(\ga_0\ga_1 F^{01}+\cdots+
           \ga_{D-2}\ga_{D-1}F^{D-2D-1} \right),
\end{eqnarray*}
Noting that $(\ga_{\mu})^2 = {\bf 1}$, i.e the
identity matrix, the defining anticommutation relations imply
\begin{eqnarray*}
 (\ga_i\ga_{i+1})^2 & = & {\bf 1}, \;\;\;\;\;\forall\;\; i=0..(D-2).
\end{eqnarray*}
Consequently we can perform an expansion of the following exponential
\begin{eqnarray*}
 \exp (i\ga_0\ga_1 F^{01}) & = & {\bf
1}+i\ga_0\ga_1F^{01}-\frac{(F^{01})^2}{2!}{\bf 1}
   -i\ga_0\ga_1\frac{(F^{01})^3}{3!} + \cdots \\
 & = & \cos F^{01}{\bf 1} + i\ga_0\ga_1 \sin F^{01},
\end{eqnarray*}
which, on consideration of the canonical representation for the tensor
$F^{\mu\nu}$, gives
\begin{eqnarray*}
 \exp (i\ga_0\ga_1 F^{01}) & = & \cos \la_1{\bf 1} + i\ga_0\ga_1 \sin \la_1 \\
 & = & \left({\bf 1}-i\ga_0\ga_1 \frac{\mbox{d}}{\mbox{d}\la_1}\right)
                     \cos \la_1.
\end{eqnarray*}
This result allows the exponential function of the full invariant matrix
$\si \cdot F$,
\begin{eqnarray*}
 \exp(\frac{1}{2}\si F) & = & \exp\left(i\ga_0\ga_1 F^{01}+\cdots+
           i\ga_{D-2}\ga_{D-1}F^{D-2D-1}\right),
\end{eqnarray*}
where incidentally all the matrices in the exponent commute with each
other, to be evaluated as
\begin{eqnarray*}
 \exp(\frac{1}{2}\si F) & = & \exp\left(i\ga_0\ga_1 F^{01}\right)
                   \exp\left(i\ga_2\ga_3 F^{23}\right)\cdots
               \exp\left(i\ga_{D-2}\ga_{D-1}F^{D-2D-1}\right) \\
 & = & \left({\bf 1}-i\ga_0\ga_1 \frac{\ptl}{\ptl\la_1}\right)\cdots
    \left({\bf 1}-i\ga_{D-2}\ga_{D-1} \frac{\ptl}{\ptl\la_{D/2}}\right)
            \left[\cos \la_1\cdots \cos \la_{D/2}\right] \\
 & = & \left[\begin{array}{ccl}
 {\bf 1} & + & \left(i\ga_0\ga_1 \frac{\ptl}{\ptl\la_1}+
                i\ga_2\ga_3 \frac{\ptl}{\ptl\la_2}+\cdots\right) \\
         & - & \left(i\ga_0\ga_1\ga_2\ga_3 \frac{\ptl}
                       {\ptl\la_1\ptl\la_2}+\cdots\right)   \\
         & + & \;\;\; \cdots \\
         & - & \left(\ga_1\ga_2\cdots\ga_{D-2}\ga_{D-1}\frac{\ptl}
                  {\ptl\la_1 \cdots\ptl\la_{D/2}}\right)
             \end{array}\right]
    \left[\cos \la_1\cdots \cos \la_{D/2}\right].
\end{eqnarray*}
Hence if we take the trace of this expression we obtain
\begin{eqnarray*}
 {\rm Tr}\; \exp(\frac{1}{2}\si F) & = & {\rm Tr}\;({\bf 1})
           \left[\cos \la_1\cdots \cos \la_{D/2}\right] \\
 & = & 2^{[D/2]} \prod_{j=1}^{D/2} \cos \la_j \\
 & = & 2^{[D/2]} \exp \left[\frac{1}{2}{\rm Tr}\; \ln \cos F \right],
\end{eqnarray*}
where the factor of $1/2$ which enters in the last line is due to the
form of the canonical representation for $F^{\mu\nu}$.
Substituting this evaluated form of the spinor trace into the
effective Lagrangian results in
\begin{eqnarray*}
 {\cal L}_I^{(1/2)} & = & -\frac{2^{[D/2]}}{2(4\pi)^{D/2}} \int_0^{\infty}
                        \frac{\mbox{d}s}{s^{1+D/2}}e^{-m^2 s}
  \exp\left[\frac{1}{2} {\rm Tr}\; \ln f(iF) \right],
\end{eqnarray*}
where
\begin{eqnarray*}
 f(z) & = & \frac{esz}{\tanh esz}.
\end{eqnarray*}
This is precisely the form required to use the techniques developed in the
scalar case. We will proceed with this process later on, although now we
should consider evaluation of the spinor trace for odd dimensional
space-times.

\subsubsection{Odd $D$}
In Euclidean space the canonical representation for the
electromagnetic field tensor is given by the even dimensional case augmented
by a zero eigenvalue $\la_0 =0$. This is a consequence of the form of the
characteristic identity for an odd-dimensional antisymmetric matrix which
implies the existence of at least one vanishing eigenvalue. If we label
the eigenvalues  $\la_0=0, \;\la_1, \;\la_2, \ldots,
\;\la_{[D/2]}$ then the analysis can proceed in an analogous manner to
the even dimensional case. i.e. we obtain
\begin{eqnarray*}
 {\rm Tr}\; \exp(\frac{1}{2}\si F) & = & {\rm Tr}\;({\bf 1})
           \left[\cos 0\cos \la_1\cdots \cos \la_{[D/2]}\right],
\end{eqnarray*}
where the extra eigenvalue has been included for clarity. Our labelling of
the eigenvalues now means that this expression is in exactly the same form
as the even dimensional case and thus the interaction Lagrangian
\begin{eqnarray}
 {\cal L}_I^{(1/2)} & = & -\frac{2^{[D/2]}}{2(4\pi)^{D/2}} \int_0^{\infty}
                        \frac{\mbox{d}s}{s^{1+D/2}}e^{-m^2 s}
  \exp\left[\frac{1}{2} {\rm Tr}\; \ln f(iF) \right], \label{l1/2}
\end{eqnarray}
where
\begin{eqnarray*}
 f(z) & = & \frac{esz}{\tanh esz},
\end{eqnarray*}
applies in a space-time of arbitrary dimension.

We may wonder at this stage as to the nature of any parity violating
terms which may appear in odd dimensional space-times.
These terms actually arise via the free
field Lagrangian and must be considered separately, as we shall
subsequently do.

\subsubsection{Evaluation of the Effective Interaction Lagrangian}
Having performed our manipulation of the Brown \& Duff spin-1/2 1-loop
Lagrangian into the form given by Eq. \ref{l1/2} we can follow our
discussion of the scalar case, extracting a
given order of the interaction (i.e. $2N$-photon scattering), and
expressing
it in the following manifestly covariant form
\begin{eqnarray*}
 {\cal L}_I^{(1/2)} & = & -\frac{2^{[D/2]}}{2(4\pi)^{D/2}} \int_0^{\infty}
                        \frac{\mbox{d}s}{s^{1+D/2}}e^{-m^2 s}
    \sum_{\{n_i\}}^{n_1+2n_2+\cdots+Nn_N=N} \frac{1}{n_1!n_2!\ldots n_N!}
              \\
  & & \;\;\; \times(a_1 {\rm Tr}\;F^2)^{n_1}(a_2 {\rm Tr}\;F^4)^{n_2}\cdots
                 (a_N {\rm Tr}\;F^{2N})^{n_N}.
\end{eqnarray*}
The technique of Delbourgo \& Matsuki \cite{delbourgo85} can now be used
to determine the coefficients $a_m$ in an analogous manner to the
calculation in the scalar case.

Following this procedure we have
\begin{eqnarray*}
 a_m & = & \frac{i(-1)^{m+1}}{8\pi m} \oint \mbox{d}z z^{-2m}
          \frac{\mbox{d}}{\mbox{d}z} \ln f(z),
\end{eqnarray*}
where in the case of the spin-1/2 Lagrangian
\begin{eqnarray*}
 f(z) & = & \frac{esz}{\tanh (esz)}.
\end{eqnarray*}
The integration in this case makes use of the following identities
\begin{eqnarray*}
 \frac{z}{e^z -1} & = & \sum_{n=0}^{\infty} \frac{B_n}{n!}z^n, \\
 \frac{z}{e^z +1} & = & \sum_{n=0}^{\infty} \frac{B_n}{n!}(1-2^n)z^n,
\end{eqnarray*}
where, again, the symbol $B_n$ represents the Bernoulli numbers.
We obtain in this case
\begin{eqnarray*}
 a_m & = & \frac{(-1)^{m+1}(2es)^{2m}}{4m} \frac{B_{2m}}{(2m)!}(2-2^{2m}).
\end{eqnarray*}

Hence, for a $2N$-photon process mediated by a virtual
spin-1/2 fermion
loop, the interaction Lagrangian is given by
\begin{eqnarray*}
 {\cal L}_I^{(1/2)} & = & -\frac{2^{[D/2]}}{2(4\pi)^{D/2}} \int_0^{\infty}
                        \frac{\mbox{d}s}{s^{1+D/2}}e^{-m^2 s}e^{2N}s^{2N}
    \sum_{\{n_i\}}^{n_1+2n_2+\cdots+Nn_N=N} \frac{1}{n_1!n_2!\ldots n_N!}\\
  & & \;\;\; \times(a'_1 {\rm Tr}\;F^2)^{n_1}(a'_2 {\rm Tr}\;F^4)^{n_2}\cdots
                 (a'_N {\rm Tr}\;F^{2N})^{n_N},
\end{eqnarray*}
where
\begin{eqnarray*}
 a'_m & = & \frac{(-1)^{m+1}2^{2m}}{4m} \frac{B_{2m}}{(2m)!}(2-2^{2m}).
\end{eqnarray*}
The final integration can be performed as in the scalar case giving
\begin{eqnarray*}
 {\cal L}_I^{(1/2)} & = & -\frac{(e^2)^N}{(4\pi)^{D/2}}
                  \frac{2^{[D/2]-1}}{(m^2)^{2N-D/2}}
      \Ga(2N-D/2)\sum_{\{n_i\}}^{n_1+2n_2+\cdots+Nn_N=N}
          \frac{1}{n_1!n_2!\ldots n_N!} \\
 & & \times(a'_1 {\rm Tr}\;F^2)^{n_1}(a'_2 {\rm Tr}\;F^4)^{n_2}\cdots(a'_N
                 {\rm Tr}\;F^{2N})^{n_N}.
\end{eqnarray*}
The expression comes of course with the
proviso that the $\Ga$-function is finite or can at least be handled
via regularisation.

We can obtain a compact form for the effective Lagrangian by noting the
following correspondance between the Bernoulli numbers and the coefficients
of the Bernoulli polynomial \cite{prudnikov},
\[ \left( 2-2^n \right) B_n = 2^n B_n\left(\frac{1}{2}\right).\]

With this substitution the coefficients are given by
\begin{eqnarray*}
 a'_m & = & \frac{(-1)^{m+1} 2^{4m}}{4m (2m)!} B_{2m}\left(\frac{1}{2}\right),
\end{eqnarray*}
and we can write the effective interaction Lagrangian in the following
form
\begin{eqnarray*}
 {\cal L}_I^{(1/2)} & = & -\frac{(e^2)^N}{(4\pi)^{D/2}}
            \frac{2^{D/2]-1}}{(m^2)^{2N-D/2}}
      \Ga(2N-D/2)
      \sum_{\{n_i\}}^{n_1+2n_2+\cdots+Nn_N=N}
          \frac{1}{n_1!n_2!\ldots n_N!} \\
 & & \times\left(2B_2 \left(\frac{1}{2}\right)\;
               {\rm Tr}\;F^2\right)^{n_1}
     \left(-\frac{4}{3}B_4\left(\frac{1}{2}\right)\;
              {\rm Tr}\;F^4\right)^{n_2} \nonumber\\
 & & \;\;\;\;\;\;\;\;\;\;\;\;\;\;	      \cdots
     \left(\frac{(-1)^{N+1}2^{4N}}{4N(2N)!} B_{2N}\left(\frac{1}{2}\right)\;
        {\rm Tr}\;F^{2N}\right)^{n_N}.
\end{eqnarray*}
Hence the full effective Lagrangian, obtained by adding in the free field
Lagrangian, is given in almost all dimensions by
\begin{eqnarray}
 {\cal L}^{(1/2)} & = & -\frac{1}{4} {\rm Tr}\; F^2 \nonumber\\
 & & -\frac{(e^2)^N}{(4\pi)^{D/2}}
            \frac{2^{D/2]-1}}{(m^2)^{2N-D/2}} \Ga(2N-D/2)
     \sum_{\{n_i\}}^{n_1+2n_2+\cdots+Nn_N=N}
          \frac{1}{n_1!n_2!\ldots n_N!} \nonumber\\
 & & \times\left(2B_2\left(\frac{1}{2}\right)\;
               {\rm Tr}\;F^2\right)^{n_1}
     \left(-\frac{4}{3}B_4\left(\frac{1}{2}\right)\;
              {\rm Tr}\;F^4\right)^{n_2} \nonumber\\
 & &  \;\;\;\;\;\;\;\;\;\;\;\;\;	      \cdots
     \left(\frac{(-1)^{N+1}2^{4N}}{4N(2N)!}B_{2N}\left(\frac{1}{2}\right)\;
        {\rm Tr}\;F^{2N}\right)^{n_N}. \label{sp1/2}
\end{eqnarray}
The qualification is placed on this statement due to the existence of
certain parity violating terms present in odd dimensional space-times which
are expected due to the presence of fermion mass terms in the classical
action.
Delbourgo \& Waites \cite{delbourgo94} have
shown that these terms have the form
\begin{eqnarray*}
 C \ep_{\mu_1\mu_2\cdots\mu_D}
         A^{\mu_1}F^{\mu_2\mu_3}\cdots F^{\mu_{D-1}\mu_D}.
\end{eqnarray*}
Due to the order of this expression in the electromagnetic field tensor
$F$, it is
apparent that this term can only relate to a $2N$-photon process such that
\begin{eqnarray*}
 D & = & 2(2N)-1.
\end{eqnarray*}
The relationship implies that in only one dimension $D$, for a given
$2N$-photon interaction, will the parity violating term emerge. i.e. for a
2 photon process $D=3$, for a 4 photon process $D=7$, etc. Interestingly,
these are the dimensions for which a charge conjugation operator exists.
Delbourgo \& Waites
have determined the general form of the coefficient $C$ which, for an
interaction of $2N$ soft photons, is given by
\begin{eqnarray*}
 C & = & \frac{e^{2N}}{2(2N)!(4\pi)^{2N-1}}.
\end{eqnarray*}
Hence, for such a process in a space-time of dimension $D=2(2N)-1$ ($N$
integral of course), the effective Lagrangian is given by
\begin{eqnarray}
 {\cal L}^{(1/2)} & = & -\frac{1}{4} {\rm Tr}\; F^2 \nonumber\\
 & & -\frac{(e^2)^N}{(4\pi)^{D/2}}
            \frac{2^{D/2]-1}}{(m^2)^{2N-D/2}}
      \Ga(2N-D/2)
      \sum_{\{n_i\}}^{n_1+2n_2+\cdots+Nn_N=N}
          \frac{1}{n_1!n_2!\ldots n_N!} \nonumber\\
 & & \times\left(2B_2\left(\frac{1}{2}\right)\;
               {\rm Tr}\;F^2\right)^{n_1}
     \left(-\frac{4}{3}B_4\left(\frac{1}{2}\right)\;
              {\rm Tr}\;F^4\right)^{n_2} \nonumber\\
 & & \;\;\;\;\;\;\;\;\;\;\;\;
	      \cdots
     \left(\frac{(-1)^{N+1}2^{4N}}{4N(2N)!}B_{2N}\left(\frac{1}{2}\right)\;
        TrF^{2N}\right)^{n_N} \nonumber\\
 & & +\frac{e^{2N}}{2(2N)!(4\pi)^{2N-1}}
          \ep_{\mu_1\mu_2\cdots\mu_D}
         A^{\mu_1}F^{\mu_2\mu_3}\cdots F^{\mu_{D-1}\mu_D}. \label{pvt}
\end{eqnarray}
In practice the only term of major importance is likely to be that of two
photon vacuum polarisation in $D=3$, the extra
invariant in this case being the Chern-Simons term. It is however
useful, for reasons of generality, to have this all encompassing result.
The form of such topological mass terms has been considered by Coleman \&
Hill \cite{coleman85} in QED$_3$ and by Reuter \& Dittrich
\cite{reuter86}, and Delbourgo \& Waites \cite{delbourgo94} in arbitrary
dimensional space-time. It
was shown by Coleman \& Hill in QED$_3$, and in arbitrary dimensions by
Delbourgo \& Waites, that the particular parity violating
term considered is indeed the only possible form allowed for reasons of
gauge invariance. This results in the strict correspondence between the
number of photons in the interaction and the dimensionality of space-time.

Returning to a more general discussion we can observe that the interaction
Lagrangian is very similar to the
expression obtained in the scalar case, with the only significant
discrepancies being the usual spinor factor $2^{[D/2]}$, the different
field invariant coefficients, and the possibility of parity violating
terms. The introduction of the Bernoulli polynomial coefficients to
describe the
coefficients of the field invariants has exhibited the following
correspondance:
\begin{eqnarray*}
 B_{2m}\left(0\right)\;\;\; & \mbox{for} &
               \mbox{spin-0 propagator particles}, \\
 2^{2m}B_{2m}\left(\frac{1}{2}\right)\;\;\; & \mbox{for} &
               \mbox{spin-1/2 propagator particles}.
\end{eqnarray*}
It is perhaps a rather large leap of faith to suggest that the
Bernoulli polynomial parameter is related to the spin of
the virtual propagator particles, but at least this does introduce a very
compact and illuminating representation.

As a particular example we can consider the case of four photon
scattering ($N=2$) in arbitrary dimensional space-time. In this case the
constraint over the $\{n_i\}$ implies that the only two terms allowed
will have $n_1=2$  and $n_2=1$. Thus we immediately recover the
standard electromagnetic field invariants
\begin{eqnarray*}
 ({\rm Tr}\;F^2)^2, \;\;\;\; \mbox{and} \;\;\;\; {\rm Tr}\;F^4.
\end{eqnarray*}
The coefficients can be evaluated easily from the general expression, and
in the case of four photon scattering we obtain
\begin{eqnarray*}
 {\cal L}_I^{(1/2)} & = & \frac{1}{720}\frac{2^{[D/2]}}{(m^2)^{4-D/2}}
     \frac{e^4}{(4\pi)^{D/2}} \Ga(4-D/2)
         \left[14 {\rm Tr}\; F^4 - 5({\rm Tr}\;F^2)^2 \right].
\end{eqnarray*}
Using a technique due to Ackhiezer \& Berestetskii \cite{ack65} this
result was verified from first principles by evaluating part of
the Feynman amplitude for four photon scattering in arbitrary dimensional
space-time. If we take the four dimensional case we obtain
\begin{eqnarray*}
 {\cal L}_I^{(1/2)}(D=4) & = & \frac{e^4}{2880\pi^2 m^4}
              \left[14 {\rm Tr}\; F^4 - 5({\rm Tr}\;F^2)^2 \right],
\end{eqnarray*}
which is the standard result obtained by Karplus \& Neuman
\cite{karplus50} and Schwinger \cite{schwinger51}.

This concludes the discussion of $2N$-photon scattering and to summarise
we have obtained expressions (Eq. \ref{sp0},\ref{sp1/2},\ref{pvt})
for the effective
Lagrangian of the electromagnetic field which can describe photon-photon
scattering to any order in arbitrary dimensions, where the virtual
propagator particles may be either spin-0 or spin-1/2 particles.

\section{2N+1-Photon Case}
The case of interactions between an odd number of photons, although not
perhaps describing pure scattering, is worthy of comment to complete the
discussion. If we consider the case of an odd number of photons
interacting via a virtual fermion loop then the case of even $D$ is trivial
and results in a vanishing interaction Lagrangian as a consequence of
Furry's Theorem.

For odd dimensions the situation is not quite so simple as Furry's
theorem no longer applies when $D=4k+1$ and there exists the possibility of
having
parity violating invariants similar in form to those which emerged in
$2N$-photon interactions. Delbourgo \& Waites \cite{delbourgo94} have
shown that these terms are given by
\begin{eqnarray*}
 \frac{e^{2N+1}}{2(2N+1)!(4\pi)^{2N}} \ep_{\mu_1\cdots\mu_D}
    A^{\mu_1}F^{\mu_2\mu_3}\cdots F^{\mu_{D-1}\mu_D},
\end{eqnarray*}
for a $2N+1$-photon process ($N$ integral). For a given $2N+1$-photon
interaction this term is only present in a
space-time with dimension $D$ given by
\begin{eqnarray*}
 D & = & 2(2N)+1,
\end{eqnarray*}
which is the set of dimensionalities for which the charge conjugation
operator does not exist.
Consequently, we can write the effective electromagnetic
field Lagrangian in this case as
\begin{eqnarray*}
 {\cal L} & = & -\frac{1}{4}{\rm Tr}\; F^2
    +\; \frac{e^{2N+1}}{2(2N+1)!(4\pi)^{2N}} \ep_{\mu_1\cdots\mu_D}
    A^{\mu_1}F^{\mu_2\mu_3}\cdots F^{\mu_{D-1}\mu_D},
\end{eqnarray*}
with the term of primary interest corresponding to a 3 photon interaction
in $5D$ when
\begin{eqnarray*}
 {\cal L} & = & -\frac{1}{4}{\rm Tr}\; F^2
  + \frac{e^{3}}{192\pi^2} \ep_{\mu\nu\rh\si\la}
    A^{\mu}F^{\nu\rh}F^{\si\la}.
\end{eqnarray*}
As mentioned earlier Coleman \& Hill \cite{coleman85}, and Delbourgo \&
Waites \cite{delbourgo94}, have shown that these are the only possible terms
which may arise, due to gauge invariance considerations, and thus
interactions of an odd number of photons in odd dimensional space-times
only have a non-zero effective interaction Lagrangian in the specific cases
described above.

\section{Relations Between Invariants}
When considering effective Lagrangians in arbitrary space-time dimensions
it turns out that the electromagnetic field invariants can suffer some
linear dependence in certain dimensionalities. This property is most
easily illustrated in terms of a particular representation for the
characteristic identities of the electromagnetic field tensor. This
representation involves writing the characteristic equation in terms of
polynomials over traces of matrix powers. A Maple routine has previously
been developed to obtain these results for any given dimension
\cite{delbourgo93}. Some examples are listed below.
\begin{eqnarray*}
\begin{array}{lc}
   D=2  & F^2 - \frac{1}{2}\;{\rm T_2}\; = 0 \\
   D=3  & F^3 - \frac{1}{2}\;{\rm T_2}\; F = 0 \\
   D=4  & F^4 - \frac{1}{2}\;{\rm T_2}\; F^2 +
        \left(\frac{1}{8} {\rm T_2^2} - \frac{1}{4} {\rm T_4}\right) = 0 \\
   D=5  & F^5 - \frac{1}{2}\;{\rm T_2}\; F^3 +
        \left(\frac{1}{8} {\rm T_2^2} - \frac{1}{4} {\rm T_4}\right)F = 0 \\
   D=6  & F^6 - \frac{1}{2}\;{\rm T_2}\; F^4 +
        \left(\frac{1}{8} {\rm T_2^2} - \frac{1}{4} {\rm T_4}\right)F^2
        -\frac{1}{48} {\rm T_2^3}+\frac{1}{8}{\rm T_2 \; T_4} - \frac{1}{6}
                  {\rm T_6} = 0,
\end{array}
\end{eqnarray*}
where ${\rm T_i}$ = Tr($F^i$).

If we take as an example a four photon scattering interaction, the two
electromagentic field invariants, ${\rm Tr}\;F^4$ and $({\rm
Tr}\;F^2)^2$, which are independent in $D=4$, are linearly
dependent in $D=2,3$. As can be seen from the characteristic identities
above, in these cases
\begin{eqnarray*}
 ({\rm Tr}\;F^2)^2 & = & 2{\rm Tr}\;F^4.
\end{eqnarray*}
Consequently the four-photon effective interaction Lagrangian simplifies
in the case of $D=2$ to
\begin{eqnarray*}
 {\cal L}_I & = & \frac{e^4}{360\pi m^6} \left({\rm Tr}\;F^2 \right)^2,
\end{eqnarray*}
and in the case of $D=3$ to
\begin{eqnarray*}
 {\cal L}_I & = & \frac{e^4}{1920\pi m^5} \left({\rm Tr}\;F^2 \right)^2.
\end{eqnarray*}

Similarly higher dimensional cases can have any linear dependence in the
invariants removed by using the relevant characteristic identity, the
general condition for linear dependence being simply that the number of
photons in the interactions is greater than the space-time dimension. This
is precisely the situation in which we would expect some linear depenence
of the photon momenta due to the limited number of degrees of freedom.
Thus it appears that the correspondence between $F^{\mu\nu}$ and the
photon momenta $k^{\mu}$, via $F^{\mu\nu} =
i(k^{\mu}A^{\nu}(k)-k^{\nu}A^{\mu}(k))$, causes any linear dependence in
the photon momenta to manifest itself as a linear dependence in the field
invariants.

It is clear that at least in the simpler cases this linear dependence can
be used to simplify the form of the effective Lagrangian. It would be
interesting to know if the corresponding linear dependence in the photon
momenta could be used to simplify calculation of the corresponding
Feynman amplitude.

\section{Conclusions}
The discussion in this paper has focussed on low-energy photon interactions
mediated by scalar and spinor propagators.  We may also consider the
possibility of interactions mediated by other forms of propagator
particles. Photon-photon scattering mediated by electroweak propagators
such as $W^{+}$ and $W^{-}$ particles has been considered by Okhlopkova
\cite{okh76}, and the gauge invariance of the amplitude in the
electroweak theory has been established by Boudjema \cite{boud87}. However,
there has apparently been no discussion of the invariant form of the
Lagrangian in the literature.
We noted earlier a possible correspondence in the
Lagrangians between the Bernoulli polynomial parameter and
the spin of the propagator particles. This relationship could be further
tested by considering the form of the effective interaction Lagrangian for a
photon scattering process mediated by a vector boson loop, i.e. $W^{\pm}$
spin-1 particles.
Taking the correspondence literally we might expect the effective
Lagrangian to have a similar form to the expressions obtained for the
scalar and spinor cases, except that the invariants would
have coefficients involving, in some manner, the constants
\begin{eqnarray*}
 B_{2m}\left(1\right).
\end{eqnarray*}

At this stage, this is obviously only pure supposition but continuing in the
same vein
we may be able to consider a slightly more believable generalisation. The
effective Lagrangians obtained by Brown \& Duff \cite{brown75} for the
scalar and spinor cases indicated that essentially the only difference in
the spinor case was the addition of an extra trace over spinor indices.
Consequently we might predict that in the vector case similar expressions
may result with the trace over spinor indices replaced by a trace of the
form
\begin{eqnarray*}
 {\rm Tr}\; \exp \left[es \Si \cdot F \right],
\end{eqnarray*}
where $\Si_{\mu}$ are the matrices equivalent to the Dirac matrices but for
spin-1 particles. Realistically there may be Fadeev-Popov ghost contributions
to
take into account here, but the form of the 1-loop Lagrangians is certainly
suggestive of the effective Lagrangian for photon scattering mediated by
any electroweak propagator particle having a similar form to the
expressions obtained in this paper.

\section{Acknowledgements}

After completion of the original preprint the authors were
made aware, for the first time, of a paper by
M.G. Schmidt and C. Schubert published in {\em Phys. Lett. B} {\bf 318}:438
(1993) in which the generating functional for the effective Lagrangian (Eq. 9)
was obtained. This result, obtained using a very different technique,
apparently coincides with ours, and thus allows us some more confidence in our
calculations. We note further that the paper of Schmidt and Schubert does not
explicitly consider the form of the Lagrangian, nor did they consider the case
of a scalar field.
The authors are also grateful to C. Schubert for pointing out an error in
the original preprint.

Finally, one of the authors, A.R., would like to acknowledge the generous
support of a
summer research studentship from the Department of Physics at the
University of Tasmania.


\begin{thebibliography}{10}

\bibitem{halp34}
O.~Halpern,
\newblock {\em Phys. Rev.} {\bf {\bf 44}}, 855 (1934).

\bibitem{eul35}
H.~Euler and B.~Kockel,
\newblock {\em Naturwiss.} {\bf {\bf 23}}, 246 (1935).

\bibitem{eul36}
H.~Euler,
\newblock {\em Ann. d. Physik} {\bf {\bf 26}}, 398 (1936).

\bibitem{heis36}
W.~Heisenberg and H.~Euler,
\newblock {\em Zeits. f. Physik} {\bf {\bf 98}}, 714 (1936).

\bibitem{karplus50}
R.~Karplus and M.~Neuman,
\newblock {\em Phys. Rev.} {\bf {\bf 80}}, 380 (1950).

\bibitem{karplus51}
R.~Karplus and M.~Neuman,
\newblock {\em Phys. Rev.} {\bf {\bf 83}}, 776 (1951).

\bibitem{beres82}
V.~Berestetskii, E.~Lifshitz, and L.~Pitaevskii,
\newblock {\em Quantum Electrodynamics},
\newblock Pergamon Press Ltd., 1982.

\bibitem{brown75}
M.~Brown and M.~Duff,
\newblock {\em Phys. Rev. D} {\bf {\bf 11}}, 2124 (1975).

\bibitem{schwinger51}
J.~Schwinger,
\newblock {\em Phys. Rev.} {\bf {\bf 82}}, 664 (1951).

\bibitem{schwinger73}
J.~Schwinger,
\newblock {\em Particles, Sources, and Fields, Vol. 2},
\newblock Addison-Wesley, 1973.

\bibitem{delbourgo85}
R.~Delbourgo and T.~Matsuki,
\newblock {\em J. Math. Phys.} {\bf {\bf 26}}, 1334 (1985).

\bibitem{alvarez83}
L.~Alvarez-Gaume and E.~Witten,
\newblock {\em Nucl. Phys. B} {\bf {\bf 234}}, 269 (1983).

\bibitem{endo85}
R.~Endo and M.~Takao,
\newblock {\em Prog. Theor. Phys.} {\bf {\bf 73}}, 803 (1985).

\bibitem{prudnikov}
A.~Prudnikov, Y.~Brychkov, and O.~Marichev,
\newblock {\em Integrals and Series, {Vol.} 3},
\newblock Gordon \& Breach Science Publishers, 1990.

\bibitem{delbourgo94}
R.~Delbourgo and A.~Waites,
\newblock {\em Aust. J. Phys.} {\bf {\bf 47}}, 465 (1994).

\bibitem{coleman85}
S.~Coleman and B.~Hill,
\newblock {\em Phys. Lett. B} {\bf {\bf 159}}, 184 (1985).

\bibitem{reuter86}
M.~Reuter and W.~Dittrich,
\newblock {\em Phys. Rev. D} {\bf {\bf 33}}, 601 (1975).

\bibitem{ack65}
A.~Ackhiezer and V.~Berestetskii,
\newblock {\em Quantum Electrodynamics},
\newblock Interscience, 1965.

\bibitem{delbourgo93}
R.~Delbourgo,
\newblock { {preprint: UTAS-PHYS-93-42}}  (1993).

\bibitem{okh76}
V.~Okhlopkova,
\newblock {\em Nucl. Phys. B} {\bf {\bf 108}}, 170 (1976).

\bibitem{boud87}
F.~Boudjema,
\newblock {\em Phys. Lett. B} {\bf {\bf 187}}, 362 (1975).

\end{thebibliography}
\end{document}